\documentclass[12pt]{article}
\usepackage[dvips]{graphicx}

\makeatletter
\@addtoreset{equation}{section}

\renewcommand{\vec}[1]{\ensuremath{\mathchoice
                     {\mbox{\boldmath$\displaystyle#1$}}
                     {\mbox{\boldmath$\textstyle#1$}}
                     {\mbox{\boldmath$\scriptstyle#1$}}
                     {\mbox{\boldmath$\scriptscriptstyle#1$}}}}

\begin{document}
\begin{titlepage}
\setcounter{page}{0}
\begin{flushright}
{\baselineskip=14pt
hep-ph 0202108 \\
Kagoshima HE-02-1\\
June, 2002

}
\end{flushright}

\vspace{1.5cm}
\begin{center}
{\baselineskip=30pt
{\Large\bf
Compressible bag model and the phase structure
}

}

\vspace{1cm}
{\large
Shigenori {\sc Kagiyama}$^1$%
\footnote{e-mail: {\tt kagiyama@sci.kagoshima-u.ac.jp}}, 
Shunichiro {\sc Kumamoto}$^1$%
\footnote{e-mail: {\tt kumamoto@cosmos01.cla.kagoshima-u.ac.jp}},
Akira {\sc Minaka}$^2$%
\footnote{e-mail: {\tt minaka@edu.kagoshima-u.ac.jp}},
Akihiro {\sc Nakamura}$^1$%
\footnote{e-mail: {\tt nakamura@sci.kagoshima-u.ac.jp}},
Kazunari {\sc Ohkura}$^1$%
\footnote{e-mail: {\tt ohkura@cosmos01.cla.kagoshima-u.ac.jp}}
and 
Shogo {\sc Yamaguchi}$^1$%
\footnote{e-mail: {\tt yamaguchi@cosmos01.cla.kagoshima-u.ac.jp}}

\bigskip
{$^1$ {\sl Department of Physics, Faculty of Science, Kagoshima University, 
Kagoshima 890-0065, Japan}}

{$^2$ {\sl Department of Physics, Faculty of Education, Kagoshima University, 
Kagoshima 890-0065, Japan}}

}
\end{center}

\vspace{1.5cm}
\begin{abstract}
\normalsize\noindent
The phase structure of hadrons and quark-gluon plasma 
is investigated by two types of equation of hadron state, namely ideal 
hadron gas model and the compressible bag model.  
It is pointed out that, while the ideal gas model produces unrealistic 
extra hadron phase, the compressible bag model gives an expected and 
reasonable phase diagram even if rich hadron spectrum is taken into account.  
\end{abstract}

\end{titlepage}

\section{Introduction}
\label{intro}
The quark-gluon plasma (QGP) has been expected to appear at 
high temperature and/or high density, due to asymptotic 
freedom of QCD.  
Indeed, not a few of the experimental results upto CERN SPS incident 
energies suggest the existence of the QGP phase \cite{proc}, and 
the recent experiment at BNL RHIC may give us cleaner and richer 
signals \cite{RHIC}.

%
%

Theoretically, in the lattice QCD calculations 
\cite{latt2000, proc2, ref1, K, ref2}, 
no phase transition is observed for the two light or two light and one 
medium light flavor cases at zero baryon density.  In both cases 
only a continuous cross over is observed for realistic quark mass values.  

On the other hand, in more phenomenological approach, state equations
of QGP and hadrons are assumed, and the transition point is determined
by Gibbs condition.   In this approach, the cross over cannot be reproduced,
and that may give rise to a reason to doubt its validity. However, the authors
believe that the gross structure of the phase diagram obtained in this approach
should still remain valid far away from the cross over region.  It should also
be noted that at a finite baryon chemical potential, not much is
known from lattice QCD because of a well known technical difficulty
which makes the Monte Carlo technique inapplicable.

The purpose of the present paper is to investigate what type of the state
equation of hadrons is preferable in the phenomenological approach,
where the discussion is devoted exclusively to gross phase structure
of hadrons and QGP, neglecting the multi-quark states.  It means that the
detail structure around phase transition or cross over region is neglected.

In the phenomenological approach, there are following problems concerning 
the gross structure of phase diagrams.  
When a simple ideal gas of nucleons is used for the equation of hadron 
state and MIT bag model is used for the equation of QGP state, hadron 
phase appears at high density and at zero temperature 
(see, e.g. \cite{CRSS}).  This is because 
\begin{equation}
p_h^0/p_q^0\to 27\qquad\hbox{as}\qquad\mu_B\to\infty,\quad(T=0) \label{diffmu}
\end{equation}
where $p_h^0$ and $p_q^0$ denote pressures of ideal nucleon phase 
and (ideal) quark phase, respectively, and $\mu_B$ is baryochemical 
potential.  
To circumvent this difficulty, volume exclusion 
effects of hadrons of van der Waals type were taken into account 
(see, e.g. \cite{CRSS, KT}, see also \cite{HR}). However, the treatments 
in \cite{CRSS, KT, HR} are thermodynamically inconsistent \cite{RGSG}, 
for example, $n\ne(\partial p/\partial\mu)_T$, 
$s\ne(\partial p/\partial T)_\mu$.  Although the formulation in 
\cite{RGSG} is thermodynamically consistent, it has still a difficulty 
that pressure diverges as $n\to 1/v_0$, where $n$ is number density and 
$v_0$ is fixed hadron volume, and acausality emerges (there is no 
relativistic rigid body) as the authors themselves pointed out.  
Thus we were led to `soft core' model, that is the 
compressible bag model \cite{cmpb, dibr}.  

In addition to the difficulty that hadron phase appears at high 
density and at zero temperature, an equation of simple ideal gas of 
hadrons again suffers from the same difficulty at another region, that is, 
at high temperature and zero baryon number density \cite{panic}.   
The argument in \cite{panic} is simple.  At $\mu_B=0$ and in high 
temperature region, the pressure $p_q$ 
of QGP phase and the pressure $p_h$ of hadron phase are given by 
\begin{equation}
p_q=g_q(1/90\pi^2)T^4-B,\qquad p_h=g_h(1/90\pi^2)T^4, \label{diffT}
\end{equation}
where $g_q$ and $g_h$ are the degeneracy factors in each phase and 
$B$ is a bag constant.  
When we consider a mixed gas of lower lying nonstrange hadrons, 
that is, $\pi$, $\rho$, $N$ and $\Delta$, the degeneracy factor $g_h$ 
becomes $3+9+(7/4)(4+16)=47$ and it exceeds 
$g_q=2\cdot 8+(7/4)4\cdot 3=37$.  Thus hadron phase appears again.  
In the compressible bag model, however, the effective degeneracy factor 
of hadron phase is greatly reduced so that QGP phase is realized at high 
temperature even if infinitely many nonstrange mesons are taken into 
account \cite{crit}.  

The authors of \cite{cmpb, dibr, crit} suggested that the compressible
bag model gives the phase diagram free from above difficulties, 
but it has not shown explicitly: 
In \cite{cmpb} the equations of state only at $T=0$ are used for 
discussing neutron stars, and in \cite{crit} used are those at $\mu_B=0$ 
for multiple production in heavy ion collisions.  
In \cite{dibr} the phase diagram with finite $T$ and $\mu_B$ is calculated, 
but rich hadron spectrum is neglected.  

The unique purpose of the present paper is, then, to show that 
the compressible bag model indeed gives the naively expected phase diagram 
in all regions.  
As far as the authors know, it is the only model that satisfies the 
followings: 
\begin{description}
 \item[(a)] It is built in thermodynamically consistent formalism.  
 \item[(b)] It is valid in very high temperature regions even if rich 
hadron spectrum is taken into account.  
 \item[(c)] It is explicitly calculable in whole regions, and it gives 
well-behaved phase diagrams, i.e. there are no extra hadron phase.  
\end{description}

Since the hadron level structure affects the phase diagram, we consider 
the following two models:
\begin{description}
\item[model I:] Hadron phase consists of $N\bar N\pi$ system while
               QGP phase consists of nonstrange quarks, their anti-quarks
               and gluons.
\item[model II:] Hadron phase consists of a system of $102$ species
                of hadrons of which masses are less than
                $10.6\,{\rm GeV}$, except for hadrons with top quark
                \cite{PDG}.
                QGP phase consists of u-, d-, s-, c-, b-quarks and
                their anti-quarks and gluons.
\end{description}
Model I is a model with a few hadrons, while model II is an example 
which includes many hadrons.

While in \cite{crit} the continuous level-density function of \cite{level} 
is used, in this paper the phase diagrams are calculated with real spectrum 
of hadrons in order to avoid the model dependence.  The effect of the cutoff
in the hadron spectrum in model II is discussed in the final section.  

In section 2, for comparison the phase structure in free point-like model 
is shown with the above two types of models, where one can explicitly see 
the difficulties explained in this section.  In section 3, the equation 
of state in the compressible bag model \cite{dibr,crit} is briefly reviewed, 
and with the resulting phase diagram it is shown that the difficulty
is indeed removed.  Final section is devoted to concluding remarks.  
\section{A difficulty in the phase structure for models of free
point-like hadrons}
\label{point}
In this section, we assume that all hadrons, quarks, and gluons are
free point-like objects and discuss their phase structure.
For this purpose, we present expressions of pressures of free point-like 
particles in order to establish notations in the first subsection.  
On the basis of these expressions, the phase diagram is 
discussed and a difficulty is pointed out in the second subsection.  
\subsection{Pressures for hadron gas and QGP}
\label{pressure}
Total pressure $p$ of a mixed gas of $N$-species of free point-like 
particles is given by 
\begin{equation}
p(T,\mu_B)=\sum_{n=1}^{N}p_n(T,\mu_n), \label{totalp}
\end{equation}
where $T$ ($\mu_B$) is the temperature (baryochemical potential) 
of the gas and $p_n$ ($\mu_n$) is the pressure (baryochemical 
potential) of the $n$-th species of particles.  
The pressure $p_n$ is given by 
\begin{eqnarray}
p_n &=& \eta_n g_n T\int{{d^3\vec{k}}\over{(2\pi)^3}}
        \log\left[1+\eta_n\exp\left(-{{E_n-\mu_n}\over T}\right)\right],
       \label{pn}\\
E_n &=& \sqrt{\vec{k}^2+m_n^2}, \label{En}\\
\mu_n &=& a_n \mu_B, \label{mun}
\end{eqnarray}
where $\eta_n$ is a statistical factor that takes the value $-1$ for 
bosons and $+1$ for fermions, respectively.  The quantities $g_n$, $a_n$, 
and $m_n$ denote the degeneracy factor, the baryon number, and the 
mass of the $n$-th species of particles, respectively.  

In low temperature region, $p_n$ is expanded as follows; 
\begin{eqnarray}
p_n 
&=& {{g_n}\over{24\pi^2}}\left[\mu_n k_{nf}^3-{3\over 2}m_n^2\mu_n k_{nf}
    +{3\over 2}m_n^4\log{\left({{\mu_n+k_{nf}}\over{m_n}}\right)}\right] \cr
&& +{{g_n}\over{12}}\mu_n k_{nf}T^2+{{7\pi^2 g_n}\over{720}}
  \left[{{3\mu_n}\over{k_{nf}}}-\left({{\mu_n}\over{k_{nf}}}\right)^3\right]
  T^4, \label{low}\\
k_{nf} &=& \sqrt{\mu_n^2-m_n^2}, \label{fermi}
\end{eqnarray}
when the $n$-th species of particles are fermions.  In the above $k_{nf}$ 
is a fermi momentum.  Eq.~(\ref{low}) is applicable for $\mu_n > m_n$.  
Otherwise the pressure $p_n$ vanishes identically.  

Now let us present expressions for pressures of hadrons and QGP.  
As for total pressure $p_h$ of a mixed gas of $N$-species of free point-like
hadrons, it is given by
\begin{equation}
p_h(T,\mu_B)=\sum_{i=1}^{N}p_{h_i}(T,\mu_i), \label{totalph}
\end{equation}
where $p_{h_i}$ follows (\ref{pn})$\sim$(\ref{fermi}).  
In the followings we consider two cases for hadronic systems; 
that is, the model I and the model II which are defined in the 
introduction.  

On the other hand, total pressure of QGP $p_q$ is given by 
\begin{equation}
p_q(T,\mu_B)=\sum_{j=1}^{N'}p_{q_j}(T,\mu_j)-B, \label{totalpq}
\end{equation}
in the bag model.  Here $p_{q_j}$ follows (\ref{pn})$\sim$(\ref{fermi}) 
and $B$ is a bag constant.  The index $j$ denotes the species of particles, 
and we consider two cases, the model I and the model II.  
\subsection{Phase structure}
\label{phasedp}
Given the expressions of pressures of hadron phase and QGP phase, 
we can consider their phase structure.  
Critical points are determined by the following Gibbs condition; 
\begin{equation}
p_h(T,\mu_B)=p_q(T,\mu_B). \label{gibbs}
\end{equation}
\begin{figure}
\begin{minipage}[t]{0.47\textwidth}
\includegraphics[width=7.5truecm]{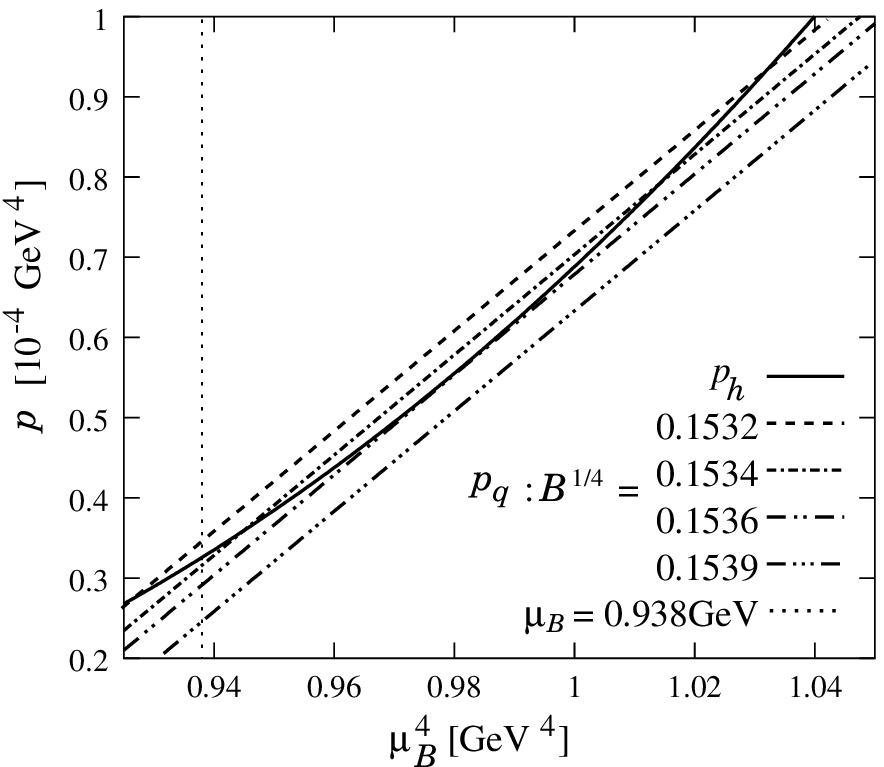}
\caption[]{$p_h$ and $p_q$ as functions of $\mu_B^4$ for the model I 
($T=0$). The unit of $B^{1/4}$ is GeV.}
\end{minipage}
\hfill
\begin{minipage}[t]{0.47\textwidth}
\includegraphics[width=7.5truecm]{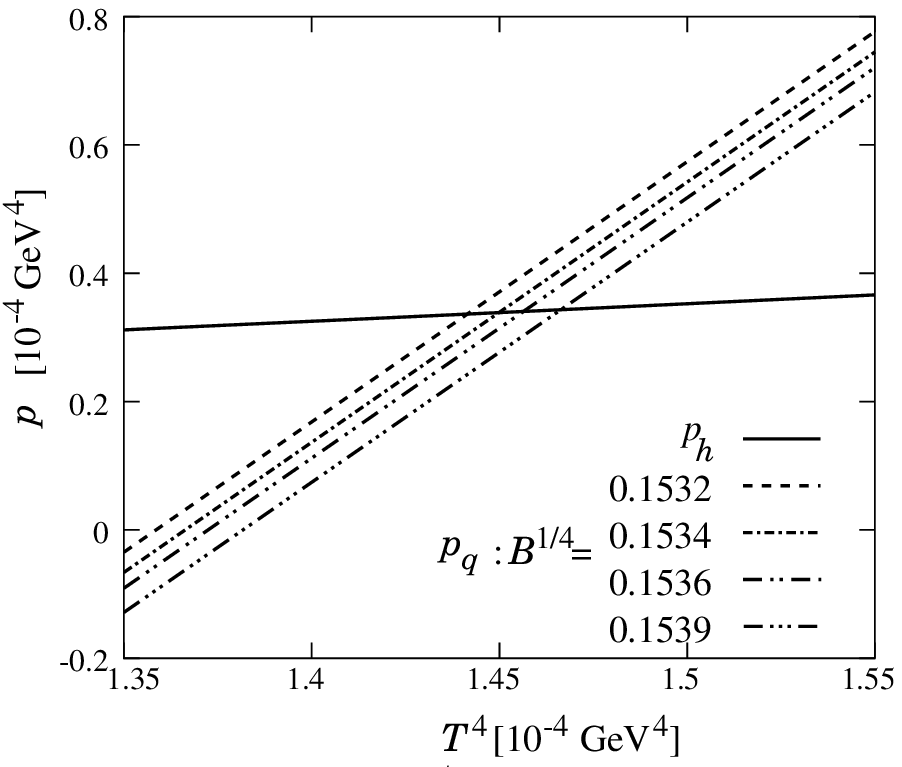}
\caption[]{$p_h$ and $p_q$ as functions of $T^4$ for the model I 
($\mu_B=0$). The unit of $B^{1/4}$ is GeV.}
\end{minipage}
\end{figure}
In Fig.~1, $p_h$ and $p_q$ are plotted as functions of $\mu_B^4$, 
where $T$ is fixed to zero.  In Fig.~2, they are plotted 
as functions of $T^4$, where $\mu_B$ is fixed to zero.  
In both figures, calculations are performed for the model I.  
As shown in Fig.~1, there is no critical point for 
$B^{1/4}>0.1536\,{\rm GeV}$.  
For $B^{1/4}=0.1536\,{\rm GeV}$, there is one critical point.  
For $0.1534\,{\rm GeV}\le B^{1/4}<0.1536\,{\rm GeV}$, 
there are two critical points.  For $B^{1/4}<0.1534\,{\rm GeV}$, 
there is one critical point since $p_h$ vanishes for $\mu_B<m_N$ 
and unphysical is an apparent critical point in the region $\mu_B<m_N$, 
where $m_N$ is nucleon mass.  Therefore, at $T=0$, 
abnormal hadron phase always appears at high densities.  

The reason for this difficulty is as follows.  In higher density 
region of $\mu_B\gg m_N$, masses are negligible so that pressures 
$p_h$ and $p_q$ are approximated as 
\begin{eqnarray}
p_h &\approx& g_h\mu_B^4, \label{pha}\\
p_q &\approx& g_q\mu_q^4, \label{pqa}
\end{eqnarray}
where $g_h$ ($g_q$) is a statistical degree of freedom of hadrons (QGP).  
Since $\mu_q=(1/3)\mu_B$, (\ref{pqa}) reads 
\begin{equation}
p_q \approx {{g_q}\over{81}}\mu_B^4. \label{pqa2}
\end{equation}
Then $p_h$ becomes large faster than $p_q$ as $\mu_B$ increases, 
since an effective statistical degree of freedom of QGP is greatly 
reduced.  In other words, hadrons get a share of energy three times 
larger than QGP, when $\mu_B$ increases, so that $p_h$ becomes 
larger than $p_q$.  As a consequence hadron phase appears in 
high density region.  

Since the difficulty mentioned above stems from the qualitative nature 
of the model of hadrons that their statistical degree of freedom is too 
large, we should consider to modify the model of hadrons.  
So far we consider the model I at present, the situation will become 
worse in the model II, since the statistical degree of 
freedom becomes further large.  

\begin{figure}
\begin{minipage}[t]{0.47\textwidth}
\includegraphics[width=7.5truecm,height=6.5truecm]{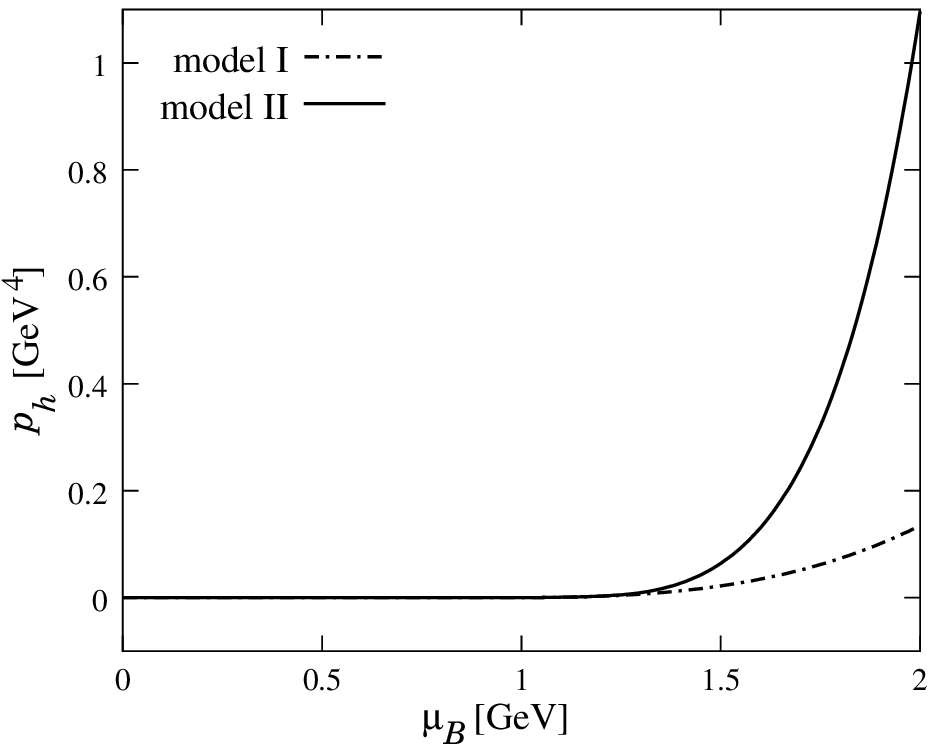}
\caption[]{$p_h$ as functions of $\mu$ for the model I and the model II 
($T=0$)}
\end{minipage}
\hfill
\begin{minipage}[t]{0.47\textwidth}
\includegraphics[width=7.5truecm]{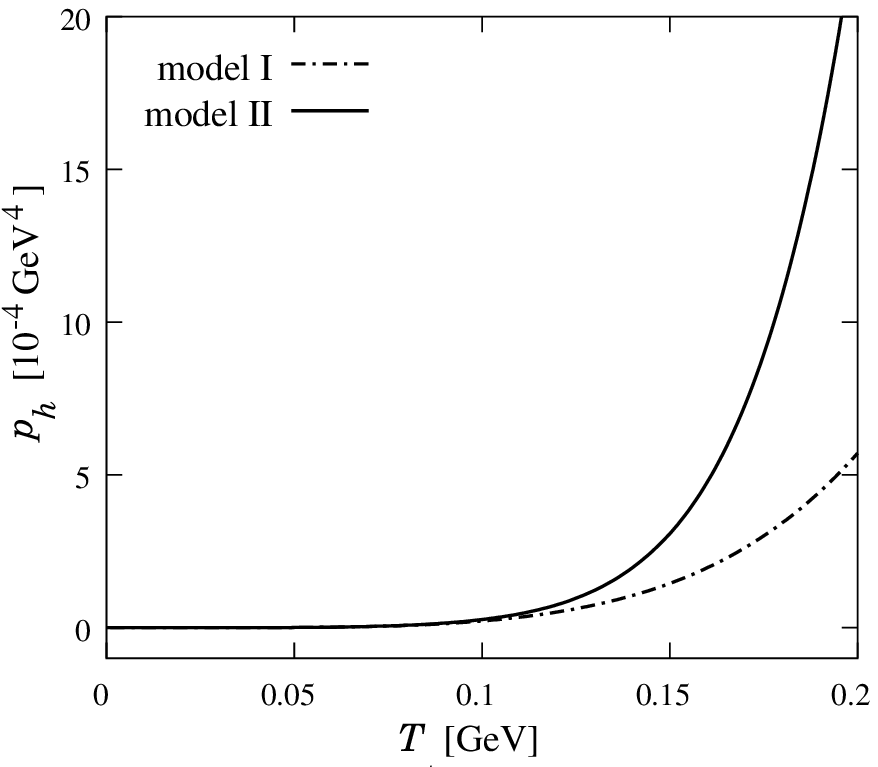}
\caption[]{$p_h$ as functions of $T$ for the model I and the model II 
($\mu_B=0$)}
\end{minipage}
\end{figure}
In Fig.~3, $p_h$'s for the model I and the model II are plotted as 
functions of $\mu_B$, where $T$ is fixed to zero and 
in Fig.~4, $p_h$'s for those are plotted as 
functions of $T$, where $\mu_B$ is fixed to zero.  
As seen in Fig.~3 (Fig.~4), $p_h$ for the model II becomes considerably 
larger than $p_h$ for the model I in high density (temperature) region, 
as expected.  

\begin{figure}
\begin{minipage}[t]{0.47\textwidth}
\includegraphics[width=7.5truecm]{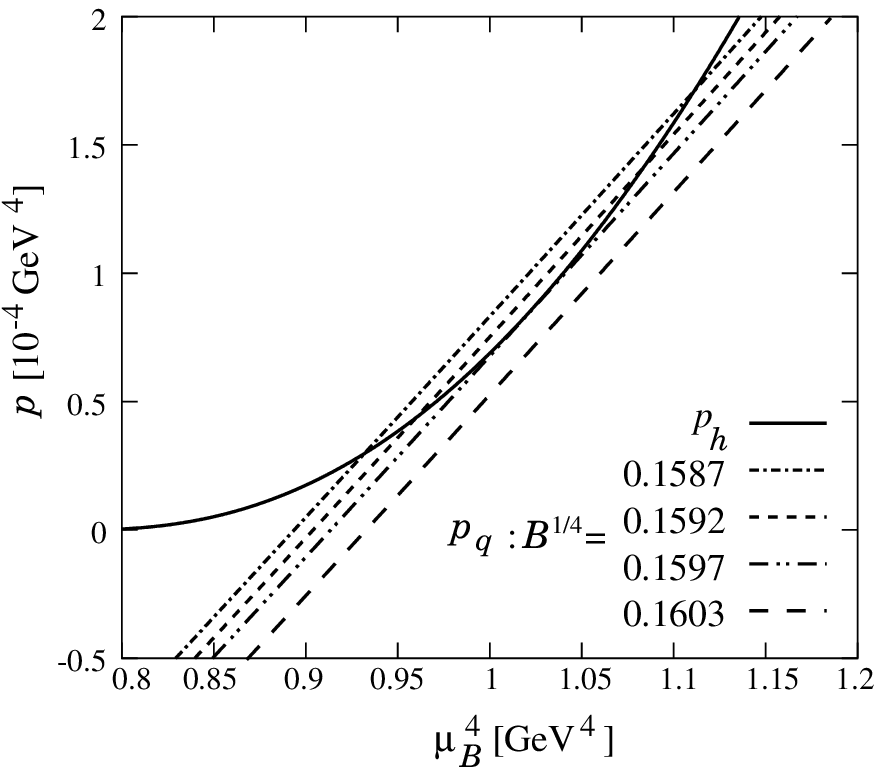}
\caption[]{$p_h$ and $p_q$ as functions of $\mu_B^4$ for the model II
($T=0$). The unit of $B^{1/4}$ is GeV.}
\end{minipage}
\hfill
\begin{minipage}[t]{0.47\textwidth}
\includegraphics[width=7.5truecm]{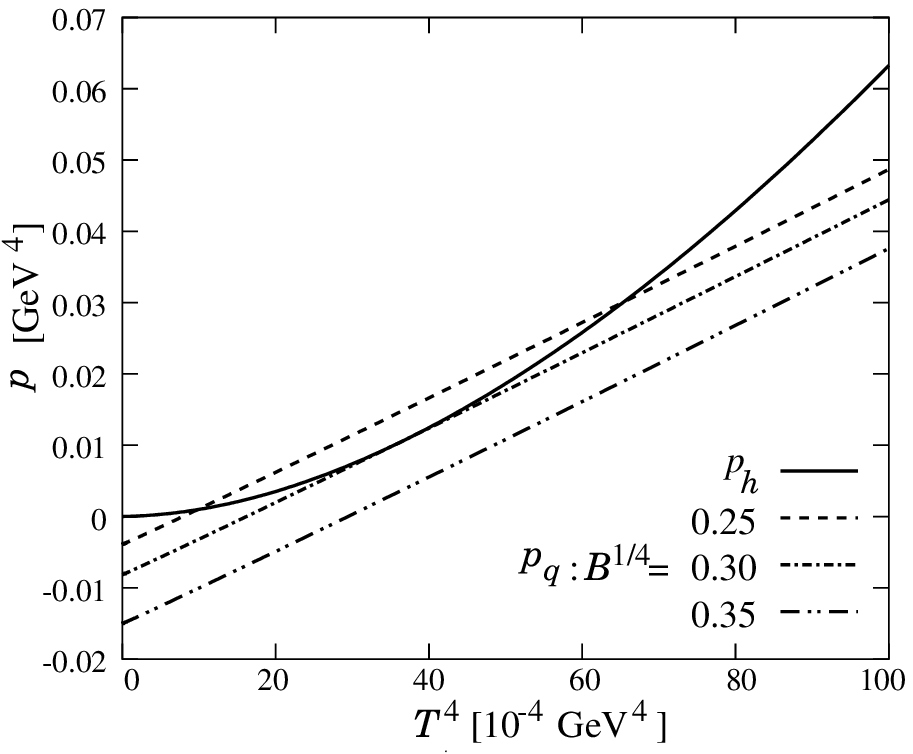}
\caption[]{$p_h$ and $p_q$ as functions of $T^4$ for the model II
($\mu_B=0$). The unit of $B^{1/4}$ is GeV.}
\end{minipage}
\end{figure}
In Fig.~5, $p_h$ and $p_q$ are plotted as functions of $\mu_B^4$ 
with $T=0$.  
In Fig.~6, they are plotted as functions of $T^4$ with $\mu_B=0$.  
In both figures, calculations are performed for the model II.  
Qualitative features in Fig.~1 and Fig.~5 are the same.  
However, qualitative features in Fig.~2 and Fig.~6 are different.  
As shown in Fig.~6, hadron phase appears in high temperature 
region.  The resulting phase diagram for the model I \cite{KT} is 
shown in Fig.~7 and that for the model II is shown in Fig.~8.  
\begin{figure}
\begin{minipage}[t]{0.47\textwidth}
\includegraphics[width=7.5truecm]{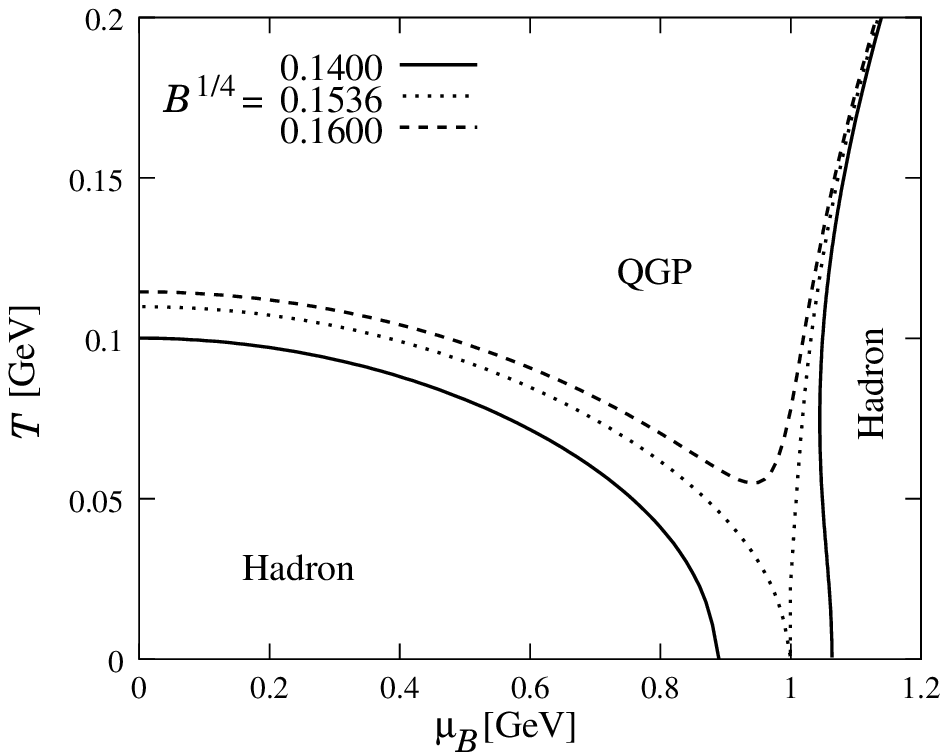}
\caption[]{Phase diagram for the model I. Besides low density and 
low temperature region, hadron phase appears in high density 
and low temperature region.  The unit of $B^{1/4}$ is GeV.}
\end{minipage}
\hfill
\begin{minipage}[t]{0.47\textwidth}
\includegraphics[width=7.5truecm,height=6.0truecm]{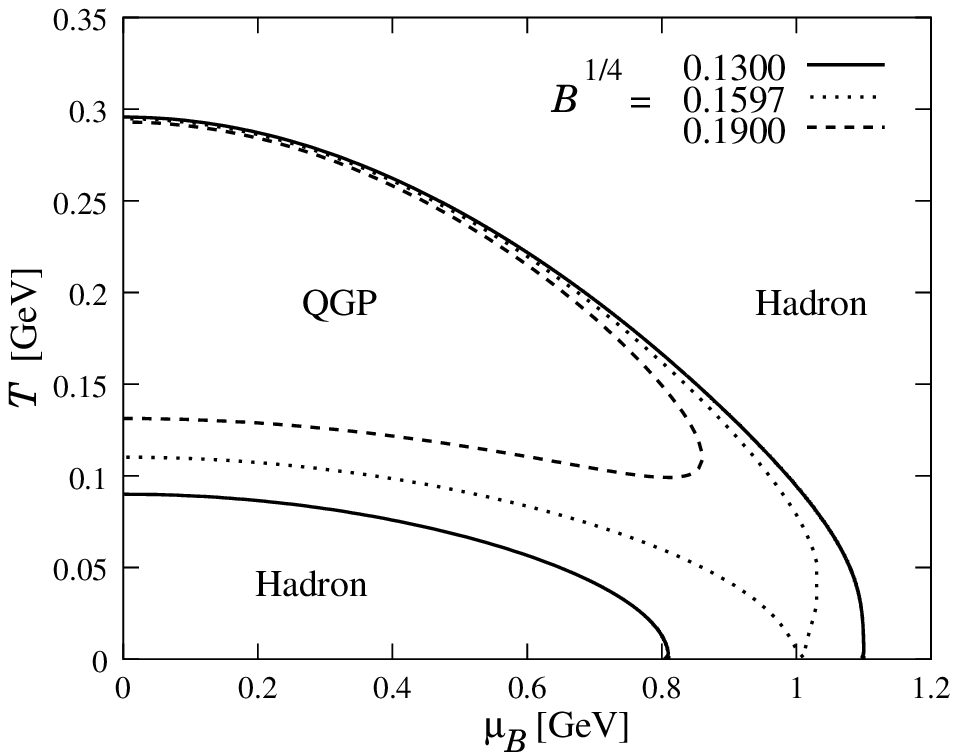}
\caption[]{Phase diagram for the model II. Besides low density and
low temperature region, hadron phase appears in high density 
and high temperature region.  QGP phase is realized only in mid 
density and mid temperature region.  The unit of $B^{1/4}$ is GeV.}
\end{minipage}
\end{figure}
The situation for the model II has become worse than for the model I.  
In order to modify this situation, we should replace the model of 
hadrons with alternative ones.  As one of them, we formulate the 
compressible bag model in the next section.  

\section{The compressible bag model}
\label{eos}
First, we briefly summarize the results of \cite{dibr, crit}.   
Let us suppose a mixed gas of $n$-species of hadrons 
enclosed in volume $V$ at finite temperature $T$.  
In the compressible bag model, the free energy function $\hat F$ 
of the gas is given by 
\begin{eqnarray}
\hat F &=& \sum_{i=1}^{n}F_f(N_i,V',T,M_i(v_i)), \label{eq:F}\\
V' &=& V-b\sum_{j=1}^{n}N_jv_j, \label{eq:Vex}
\end{eqnarray}
where $N_i$ is a number of $i$-th species of hadrons and $v_i$ is its volume 
and $M_i$ its mass.  The constant $b$ is a volume exclusion efficiency 
parameter.   The function $F_f$ is a free energy function of free 
point-like hadron gas.  As for mass function $M_i(v_i)$, we assume that 
of MIT bag model:  
\begin{eqnarray}
M_i(v_i) = A_iv_i^{-1/3}+Bv_i, \label{eq:MIT}
\end{eqnarray}
where $B$ is bag constant.  

Under the approximation that the average of inverse Lorentz factor
of $i$-th species of hadrons equals to unity ($\langle\gamma_i^{-1}
\rangle\approx 1$) in the rest frame of the system, 
basic requirement of the compressible bag model 
that $\partial\hat F/\partial v_i=0$ and the requirement that 
the chemical potential of hadron should be, 
$\mu_i=\partial\hat F/\partial N_i=a_i\mu_B$, where $a_i$ is the 
baryon number of the $i$-th species of hadrons, determine the pressure $p$ 
of the system as a function of $T$ and $\mu_B$ by the following equation: 
\begin{eqnarray}
p &=& \sum_{i=1}^{n}\eta_i g_iT\int{{d^3\vec{k}}\over{(2\pi)^3}}
\log\left\{1+\eta_i\exp[-(E_i-\mu'_i(p,m_i))/T]\right\}, \label{eq:pi}\\
&&E_i=\sqrt{\vec{k}^2+M_i(p,m_i)^2},\cr
M_i(p,m_i) &=& m_i\left(1+{{3bp}\over{4B}}\right)\left(1+{{bp}\over{B}}
\right)^{-3/4}, \label{eq:Mi}\\
\mu'_i(p,m_i) &=& \mu_i-bv_ip  
= a_i\mu_B-{{bm_ip}\over{4B}}\left(1+{{bp}\over{B}}\right)^{-3/4}, 
\label{eq:vi}
\end{eqnarray}
where $g_i$ is a degeneracy factor of the $i$-th species of hadrons and 
$m_i=4(A_i/3)^{3/4}B^{1/4}$ is its mass in the vacuum.  

Here two comments are in order.  
First $p$ determined as above depends on $b$ and $B$ only in the 
combination of $b/B$ since $M_i$ and $\mu'_i$ depend on $b$ and 
$B$ in that combination as seen in (\ref{eq:Mi}) and (\ref{eq:vi}).  
Second the pressure in the compressible bag model does not become 
so large in high density and/or high temperature region in contrast 
to the one in the model of free point-like hadrons.  This is because 
masses of hadrons become large in high density and/or high temperature 
region in the compressible bag model so that kinetic energies of hadrons 
increase rather slowly and the pressure of hadrons do not become so 
large.   Thus the compressible bag model have a chance to evade 
the difficulty discussed in the previous section.  
In the following, we assure that the difficulty is indeed removed, 
by numerical calculations.  
\begin{figure}
\begin{minipage}[t]{0.47\textwidth}
\includegraphics[width=7.5truecm]{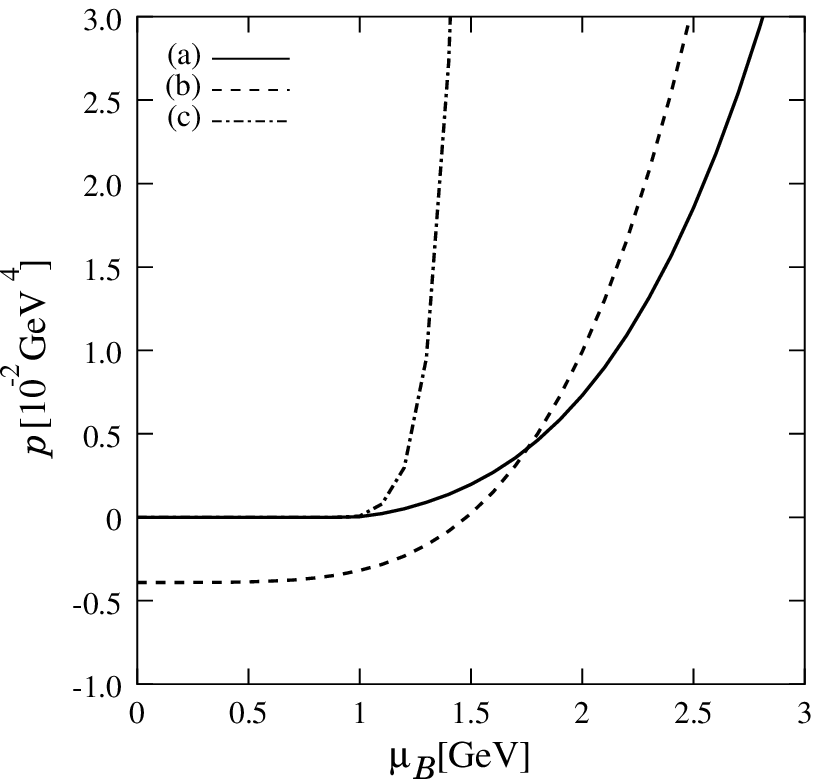}
\caption[]{(a) A solid line shows $p_h$ as a function of $\mu_B$
for the compressible bag model.  (b) A dashed line shows $p_q$ for QGP phase.  
(c) A dash-dotted line shows $p_h$ for the free point-like hadron model.
All lines are calculated for the model II with $B^{1/4}=0.25\,{\rm GeV}$ 
and $T=0$.}
\end{minipage}
\hfill
\begin{minipage}[t]{0.47\textwidth}
\includegraphics[width=7.5truecm]{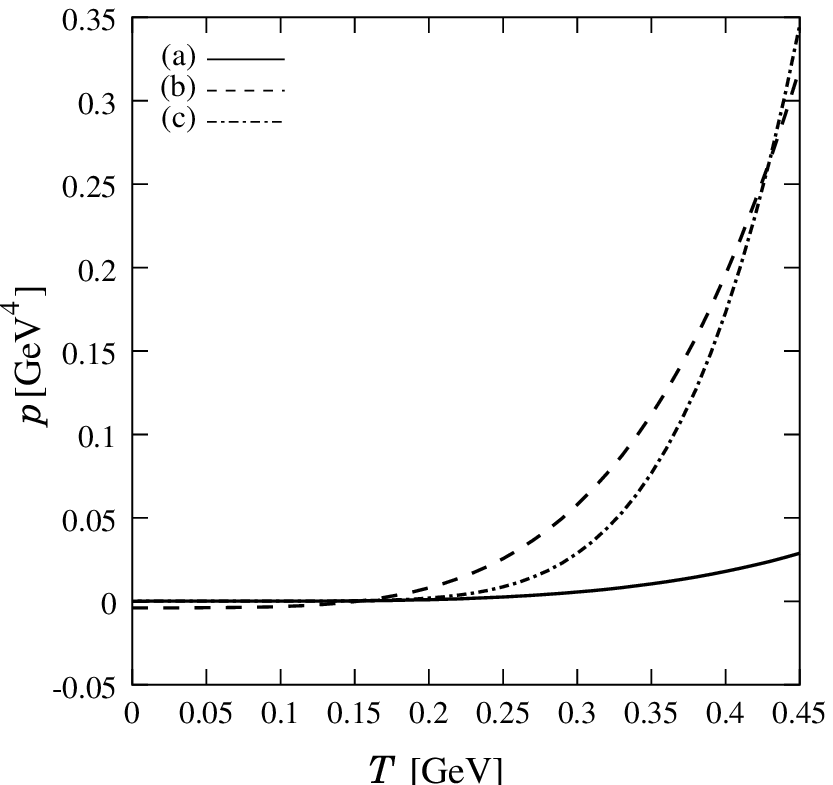}
\caption[]{(a) A solid line shows $p_h$ as a function of $T$
for the compressible bag model.
(b) A dashed line shows $p_q$ for QGP phase.  
(c) A dash-dotted line shows $p_h$ for the free point-like hadron model.
All lines are calculated for the model II with $B^{1/4}=0.25\,{\rm GeV}$ 
and $\mu_B=0$.}
\end{minipage}
\end{figure}
\begin{figure}
\begin{center}
\includegraphics[width=7.5truecm]{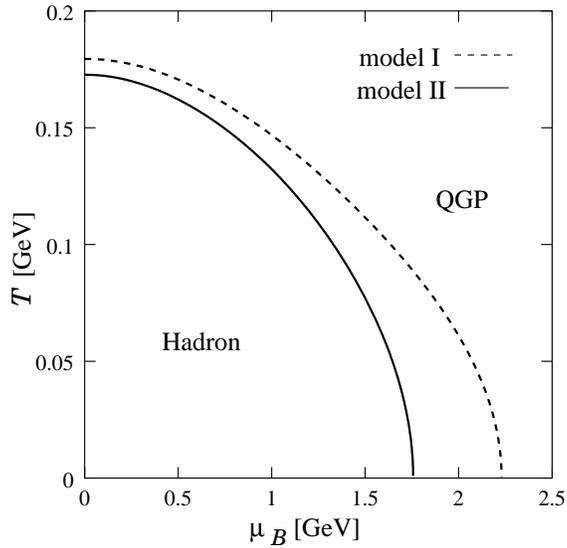}
\caption[]{Phase diagram for the model II (solid line).  For comparison,
phase diagram is shown for the model I (dashed line). Calculations are
done with $B^{1/4}=0.25\,{\rm GeV}$.}
\end{center}
\end{figure}

In order to do numerical calculations, we have to fix bag constant 
$B$ and volume exclusion parameter $b$.  
The parameter $b$ is determined by the relation 
\begin{eqnarray}
bv_N(p=0) = {{bm_N}\over{4B}} = {{4\pi}\over 3}(0.82\,{\rm fm})^3,\label{eq:Rc}
\end{eqnarray}
when the bag constant is given.  The value $0.82\,{\rm fm}$ is the 
proton charge radius.  It should be noted that equations of state 
for hadron phase are determined uniquely at this stage since the 
equation of states for hadron phase only depend on $b/B$.  
We tentatively put 
\begin{equation}
B^{1/4}=0.25\,{\rm GeV}
\end{equation}
and then get $b = 5.0$.
The resulting $p_h$ and $p_q$ are shown as functions of $\mu_B$ ($T$)
in Fig.~9 (Fig.~10).  We can recognize that the compressible bag model
does not suffer from the difficulty that exist for free point-like hadrons.
This point is more clearly seen by the phase diagrams shown in Fig.~11
with $B^{1/4}=0.25\,{\rm GeV}$.  Critical temperature $T_c$ and critical
baryochemical potential $\mu_c$ are given by
\begin{eqnarray}
&&T_c=0.18\,{\rm GeV},\qquad \mu_c=2.2\,{\rm GeV}\qquad\hbox{in model I},\\
&&T_c=0.17\,{\rm GeV},\qquad \mu_c=1.8\,{\rm GeV}\qquad\hbox{in model II}.
\end{eqnarray}

It is noted that the state equation of hadrons are determined by the ratio 
$b/B$ and that of QGP by $B$, the critical values, then, depends on $B$ 
and $b$.    If the constraint (\ref{eq:Rc}) is taken, the values depend 
only on $B$.  As a reference, some cases are shown below; 
\begin{eqnarray}
&&T_c=0.14\,{\rm GeV},\qquad \mu_c=1.7\,{\rm GeV}\qquad\hbox{in model I},\\
&&T_c=0.14\,{\rm GeV},\qquad \mu_c=1.4\,{\rm GeV}\qquad\hbox{in model II},
\end{eqnarray}
for $B^{1/4}=0.20\,{\rm GeV}$ and 
\begin{eqnarray}
&&T_c=0.22\,{\rm GeV},\qquad \mu_c=2.7\,{\rm GeV}\qquad\hbox{in model I},\\
&&T_c=0.21\,{\rm GeV},\qquad \mu_c=2.1\,{\rm GeV}\qquad\hbox{in model II},
\end{eqnarray}
for $B^{1/4}=0.30\,{\rm GeV}$.  

From Figs.~9$\sim$11 one can see that the compressible bag model gives an 
expected and reasonable phase diagram in whole regions. This result is 
stable in the sense that one can choose the bag constant from rather 
wide range.  
\section{Concluding remarks}
\label{summary}
In this paper, it has been shown by explicit calculations that the 
compressible bag model gives well-behaved phase diagram in whole regions, 
even if many hadron states are taken into account.  
In the present calculation it is true that, even in the model II, 
the infinite series of hadrons are cut off at finite mass, 
but the compressible bag model may give an expected phase diagram 
if infinite series of hadrons are included.  
This is because, in the model, the masses of hadrons increase and the 
effect of higher mass state is much suppressed in high-temperature 
or high-density regions.
Indeed in \cite{crit} a continuous level-density function are used 
for hadrons and it is shown that the abnormal hadron phase does not appear 
in the compressible bag model, although limited at $\mu_B=0$.  

Free point-like models produce unnatural hadron state in high-temperature 
or high-density regions as shown in Sections 1 and 2.  
A way to avoid this difficulty is the compressible bag model, and it is 
worth examining in various phenomenological analyses.

%

\end{document}